\documentclass[prd,nofootinbib,preprint,superscriptaddress,twocolumn,10pt]{revtex4}
\pdfoutput=1
\usepackage[utf8]{inputenc}
\usepackage{amsmath,amssymb}
\usepackage{epsfig}
\usepackage{comment}
\usepackage{bbm}
\usepackage{graphicx}
\usepackage[usenames,dvipsnames]{color}
\usepackage{float}
\usepackage{bbold}
\usepackage[colorlinks,citecolor=blue]{hyperref}
\usepackage[normalem]{ulem}

\usepackage{color}
\usepackage{subfigure}
\begin{document}
%opening
\title{Gravitational waves from seesaw assisted collapsing domain walls}

%%%%%%%%%   Authors   %%%%%%%%%%%%

\author{Debasish Borah}
\email{dborah@iitg.ac.in}
\affiliation{Department of Physics, Indian Institute of Technology Guwahati, Assam 781039, India}

\author{Indrajit Saha}
\email{s.indrajit@iitg.ac.in}
\affiliation{Department of Physics, Indian Institute of Technology Guwahati, Assam 781039, India}

\begin{abstract}
Spontaneous breaking of discrete symmetries like $Z_2$ leads to the formation of stable topological defects such as domain walls which, if allowed to dominate, can potentially be in conflict with cosmological observations. Incorporating explicit $Z_2$-breaking bias terms can lead to annihilation of such walls while also emitting stochastic gravitational waves (GW). We study the role of heavy right-handed neutrinos present in the type-I seesaw framework to generate such a bias term via quantum corrections. This offers interesting correlations among the seesaw scale, GW peak amplitude and peak frequency which can be probed at present and future experiments related to GW as well as precision measurements of the cosmic microwave background (CMB). In flavor symmetric UV complete scenarios with degenerate RHNs at leading order, such tiny coupling of RHNs to a $Z_2$-odd scalar can also lead to small mass splittings suitable for explaining the observed baryon asymmetry of the universe via resonant leptogenesis.
\end{abstract}

\maketitle
%\flushbottom

\section{Introduction}
The observation of neutrino oscillations across different experiments has established the presence of sub-eV scale neutrino mass and large leptonic mixing \cite{ParticleDataGroup:2024cfk}. As the Standard Model (SM) of particle physics can not accommodate non-zero neutrino mass and mixing, several Beyond Standard Model (BSM) proposals have been proposed to address the same in the last few decades. The simplest among them is the type-I seesaw mechanism \cite{Minkowski:1977sc, GellMann:1980vs, Mohapatra:1979ia,Sawada:1979dis,Yanagida:1980xy, Schechter:1980gr} where the SM is extended by at least two copies of heavy right-handed neutrino (RHN). Another appealing feature of the seesaw mechanism is the explanation of the baryon asymmetry of the Universe (BAU) \cite{ParticleDataGroup:2024cfk, Planck:2018vyg} via leptogenesis \cite{Fukugita:1986hr}. The BAU is another observed phenomena which the SM fails to address due to its inability to satisfy the Sakharov's conditions \cite{Sakharov:1967dj} in required amount. In type-I seesaw implementation of leptogenesis, out-of-equilibrium decay of the heavy RHNs generate a non-zero lepton asymmetry first which gets converted into BAU via electroweak sphalerons \cite{Kuzmin:1985mm}. While seesaw models have these appealing features, they are often difficult to probe at experiments. Canonical seesaw models explaining neutrino data and BAU via leptogenesis typically operate at a very high scale, away from direct experimental reach in foreseeable future. For example, successful leptogenesis with hierarchical RHNs requires the seesaw scale to be $\gtrsim 10^9$ GeV, known as the Davidson-Ibarra bound \cite{Davidson:2002qv}. While TeV scale leptogenesis is possible with quasi-degenerate RHNs \cite{Pilaftsis:2003gt}, they often require tiny Yukawa couplings. However, in the presence of additional approximate symmetries or textures, it is possible to have larger Yukawa couplings \cite{Shaposhnikov:2006nn, Kersten:2007vk, Moffat:2017feq}, which keeps such scenarios within reach of future experiments \cite{Klaric:2020phc, Drewes:2021nqr, Hernandez:2022ivz}.
As such direct detection prospects remain absent for high scale seesaw, several recent attempts also considered the possibility of probing high scale seesaw indirectly via stochastic gravitational wave (GW) observations \cite{Dror:2019syi, Blasi:2020wpy, Fornal:2020esl, Samanta:2020cdk, Barman:2022yos, Huang:2022vkf, Dasgupta:2022isg, Okada:2018xdh, Hasegawa:2019amx, Borah:2022cdx, Borah:2022vsu, Barman:2023fad, Borah:2023saq}. In these works, the sources of stochastic GW were considered to be cosmic strings \cite{Dror:2019syi, Blasi:2020wpy, Fornal:2020esl, Samanta:2020cdk, Borah:2022vsu}, domain walls (DW) \cite{Barman:2022yos, Barman:2023fad,Saikawa:2017hiv,Roshan:2024qnv,Bhattacharya:2023kws,Blasi:2022ayo,Blasi:2023sej, Borah:2024kfn, Borboruah:2024lli} or bubbles generated at first order phase transition \cite{Huang:2022vkf, Dasgupta:2022isg, Okada:2018xdh, Hasegawa:2019amx, Borah:2022cdx, Borah:2023saq}.

Motivated by these, in this work, we consider the possibility of generating stochastic GW from domain wall annihilation aided by RHNs in type-I seesaw. A $Z_2$-odd singlet scalar $\phi$ driving spontaneous symmetry breaking leads to the formation of domain walls. Typically, a bias term in the potential \cite{Zeldovich:1974uw, Vilenkin:1981zs, Sikivie:1982qv, Gelmini:1988sf, Larsson:1996sp} is introduced to get rid of such walls. Such bias terms can also arise from higher dimensional operators suppressed by the scale of quantum gravity (QG) \cite{Rai:1992xw, Lew:1993yt} or from radiative corrections \cite{Zhang:2023nrs, Zeng:2025zjp}. We consider the coupling of $\phi$ to RHNs as the origin of the bias term via radiative corrections and study the impact of the seesaw scale on DW annihilation and resulting GW amplitude as well as peak frequency. We constrain the scales of seesaw and $Z_2$-breaking for different relative contributions of $\phi$ to RHN masses and show that the low as well as intermediate scale seesaw scenarios remain within reach of several present and future GW experiments. Such low scale seesaw can also generate the observed BAU via resonant leptogenesis. While we consider a minimal setup in this work, it is possible to have UV completions where $Z_2$-breaking couplings of RHNs to $\phi$ can lead to tiny mass-splitting among RHNs as required from resonant leptogenesis point of view.

This paper is organised as follows. In section \ref{sec1}, we discuss the minimal setup followed by the details of GW and leptogenesis in section \ref{sec2} and \ref{sec3} respectively. We finally conclude in section \ref{sec4}.

\section{A Minimal Setup}
\label{sec1}
As mentioned earlier, we consider a minimal type-I seesaw framework augmented by a $Z_2$-odd singlet scalar $\phi$. Therefore, our setup consists of only three BSM fields namely, two RHNs and a real singlet scalar. The relevant part of the fermionic Lagrangian involving these new fields can be written as   
\begin{align}
    -\mathcal{L} \supset h_{\alpha i} \overline{L_\alpha} \tilde{H} N_i+ \frac{1}{2} M_i \overline{N^c}_i N_i + \frac{1}{2} y_i \phi \overline{N^c}_i N_i + {\rm h.c.} \nonumber 
\end{align}
where i=1,2 correspond to two generations of RHNs. The Yukawa coupling of $\phi$ with RHN breaks $Z_2$ symmetry explicitly playing a crucial role in domain wall annihilation as we discuss in upcoming sections. The $Z_2$-invariant scalar potential for $\phi$ is
\begin{equation}
    V(\phi) = -\frac{\mu^2_\phi}{2} \phi^2 + \frac{\lambda}{4} \phi^4
\end{equation}
where we have ignored its coupling with the SM Higgs $H$ for simplicity. At tree level, the $Z_2$ symmetry is spontaneously broken by the non-zero VEV of $\phi$, leading to degenerate minima at $\langle \phi \rangle=\pm v_\phi$. Considering $\langle \phi \rangle=- v_\phi$ to be the true minima, the mass matrix of the RHNs can be written as
\begin{equation}
    m_N = M - y v_\phi.
\end{equation}
The SM Higgs doublet $H$ is parameterized as
\begin{align}
    H=\frac{1}{\sqrt{2}} \begin{pmatrix}
        0 \\
        h +v
    \end{pmatrix},
\end{align}
with $v=246$ GeV being the vacuum expectation value (VEV). After electroweak symmetry breaking, neutrinos acquire a Dirac mass term $(M_D)_{\alpha i}= \frac{1}{\sqrt{2}} v h_{\alpha i}$. The type-I seesaw contribution to light neutrino mass in the seesaw limit $M_D \ll m_N $ can be written as
\begin{equation}
    M_\nu \simeq -M_D m^{-1}_N M^T_D\, , \quad {\rm or},\quad (M_\nu)_{\alpha\beta} \simeq -\frac{v^2}{2} \frac{h_{\alpha i} h_{\beta i}}{m_{N_i}}.
    \end{equation}
The constraints on the model parameters from neutrino oscillation data can be incorporated by the use of Casas-Ibarra (CI) parametrisation~\cite{Casas:2001sr} for type-I seesaw given by
\begin{equation}
    h_{\alpha i} =\dfrac{1}{v} (U D_{\sqrt{M_{\nu}}} R^{\dagger} D_{\sqrt{m_N}}),
    \label{eq:CI}
\end{equation}
where $R$ is an arbitrary complex orthogonal matrix satisfying $RR^{T}=\mathbbm{1}$ and $U$ is the Pontecorvo-Maki-Nakagawa-Sakata (PMNS) leptonic mixing matrix which also diagonalises the light neutrino mass matrix in a diagonal charged lepton basis. For two RHNs, $D_{\sqrt{m_N}}={\rm diag}(\sqrt{m_{N_1}},\sqrt{m_{N_2}})$. The equivalent diagonal matrix for light neutrinos is $D_{\sqrt{M_{\nu}}}={\rm diag}(m_1, m_2, m_3)$. Since there are only two RHNs, the lightest active neutrino remains massless, which is consistent with the current neutrino data. The orthogonal matrix $R$ for two RHNs is given by~\cite{Ibarra:2003up}
\begin{equation}
    R = \begin{pmatrix}
        0 & \cos{z_1} & \pm \sin{z_1} \\
        0 & -\sin{z_1} & \pm \cos{z_1}
    \end{pmatrix}
\end{equation}
where $z_1=a+ib$ is a complex angle. The diagonal light neutrino mass matrix can be written as 
\begin{align}
    D_{\sqrt{M_\nu}} = \left\{ \begin{array}{cc}
    {\rm diag}\left(
    0,\sqrt{\Delta m^2_{\rm sol}}\, , \sqrt{\Delta m^2_{\rm sol}+\Delta m^2_{\rm atm}}\right) & {\rm (NH)} \\
    {\rm diag}\left(
    \sqrt{\Delta m^2_{\rm atm}-\Delta m^2_{\rm sol}}\, ,\sqrt{\Delta m^2_{\rm atm}}\, , 0\right) & {\rm (IH)}
    \end{array}
    \right. \nonumber 
\end{align}
for normal hierarchy (NH) and inverted hierarchy (IH), respectively. The PMNS parameters as well as solar and atmospheric mass splittings can be obtained from neutrino global fit data \cite{Esteban:2024eli}.

\section{Domain walls and GW}
\label{sec2}
Spontaneous breaking of discrete symmetries like $Z_2$ can lead to the formation of topological defects like domain walls \cite{Zeldovich:1974uw, Kibble:1976sj, Vilenkin:1981zs,Saikawa:2017hiv,Roshan:2024qnv}. Since the energy density of DW redshifts inversely with the scale factor $(a^{-1})$ which is slower compared to the redshift of matter $(a^{-3})$ and radiation $(a^{-4})$, they can dominate the Universe, altering the successful predictions of standard cosmology like cosmic microwave background (CMB), big bang nucleosynthesis (BBN) etc. Assuming the walls to be formed after inflation, the simplest way to make them disappear is to introduce a small pressure difference \cite{Zeldovich:1974uw, Vilenkin:1981zs, Sikivie:1982qv, Gelmini:1988sf, Larsson:1996sp}, also known as the bias. Such bias terms for $Z_2$ domain walls are written in terms of odd powers of $\phi$, which break $Z_2$ explicitly. We consider the heavy RHNs to be responsible for generating such terms after taking quantum corrections into account \cite{Zhang:2023nrs}.

At tree level, the scalar potential for the $Z_2$-odd scalar $\phi$ has degenerate minima at $\langle \phi \rangle=\pm v_\phi$. At one-loop level, the Coleman-Weinberg correction \cite{Coleman:1973jx} can lift one of these minima as the scalar potential receives different corrections at these two minima due to RHNs inside the loop. Starting with the Yukawa Lagrangian described earlier, the field dependent masses of RHNs can be written as
\begin{align}
    m_{N_i} (\phi) = M_i + y_i \phi = M_i + \delta M_i.
\end{align}
This leads to the following one-loop correction to the scalar potential of $\phi$ at $\langle \phi \rangle = +v_\phi$
\begin{align}
    V_{\rm cw}(+v_\phi)&  = \Sigma_i \frac{-1}{64\pi^2}n_i(m_{N_i}(+v_\phi))^4 \nonumber \\  
   & \times \{{\rm Log}(\frac{(m_{N_i}(+v_\phi))^2}{\mu^2}) -3/2\}
\end{align}
where $n_i=2$ is the degrees of freedom of $N_i$ and $\mu$ is the renormalisation scale, taken to be $v_\phi$. Without any loss of generality, we consider $\langle \phi \rangle=-v_\phi$ to be the true minima such that the one-loop bias term at $T=0$ can be written as 
\begin{align}
    \Delta V_{\rm cw}= V_{\rm cw}(+v_\phi) - V_{\rm cw} (-v_\phi).
\end{align}
If the field $\phi$ couples to the bath particles at temperature $T$ which breaks $Z_2$ symmetry, it is also possible to generate a temperature dependent bias term \cite{Zeng:2025zjp}. The finite-temperature correction \cite{Dolan:1973qd,Quiros:1999jp} to the potential at the minima $\langle \phi \rangle = +v_\phi$ can be written as
\begin{align}
    V_{\rm T}(+v_\phi) & = \Sigma_i \frac{-T^4}{2\pi^2}n_i J_F \Big(\frac{(m_{N_i}(+v_\phi))^2}{T^2}\Big)\\ & \approx \Sigma_i \frac{-T^4}{2\pi^2}n_i \Big(-\frac{\pi^2}{24}\frac{(m_{N_i}(+v_\phi))^2}{T^2}\Big)\\ & = \Sigma_i \frac{n_i}{48}(m_{N_i}(+v_\phi))^2 T^2.
\end{align}
Therefore, the temperature dependent bias term is
\begin{align}
    \Delta V_{\rm T}= V_{\rm T}(+v_\phi) - V_{\rm T}(-v_\phi) = c_1 T^2.
\end{align}
For two generations of RHNs, the field dependent masses are
\begin{align}
    m_{N_1}(\phi)=M_1 + y \phi   \quad \quad m_{N_2}(\phi)=M_2 +2y\phi
\end{align}
where we assume $y_2=2y_1 = 2y$ with degenerate bare masses $M_1=M_2$. For these choices of masses, the $c_1=\frac{1}{2}yM_1v_\phi$.
The total bias term at finite temperature can be expressed as 
\begin{align}
    V_{\rm bias}(T)=\Delta V_{\rm cw} + \Delta V_{\rm T} (T).
\end{align}
For the parameter space of our interest, we find $\Delta V_{\rm cw}$ to dominate the bias term at $T=T_{\rm ann}$. When the tension force $ p_T = \sigma / L$ (where $\sigma \sim \sqrt{\lambda} v^3_{\phi}$ is the wall tension and $L \sim (\sqrt{\lambda} v_\phi)^{-1}$ is the wall size) becomes comparable to the pressure difference  $p_V = V_{\rm bias}$ \cite{Bai:2023cqj}, then the walls annihilate. This is true in the absence of friction effects between the domain walls and the thermal plasma \cite{Nakayama:2016gxi, Galtsov:2017udh, Blasi:2022ayo}.  For scalar field having tiny couplings with the bath particles, such friction effects can be ignored \cite{Babichev:2021uvl}. We keep $\lambda=0.1$ in our numerical analysis. The annihilation temperature of domain walls in our model can be found as
\begin{widetext}
\begin{align}
    T_{\rm ann} = \frac{3.39\times10^{-2} (C_{\rm ann}\mathcal{A})^{-1/2} \left (\frac{g_*}{10} \right)^{-1/4} \left (\frac{\sigma}{10^9 {\rm GeV^3}}\right )^{-1/2} \left (\frac{\Delta V_{\rm cw}}{10^{-12} {\rm GeV^4}} \right )^{1/2}}{\sqrt{1-1.14\times10^{-3} (C_{\rm ann}\mathcal{A})^{-1} \left (\frac{g_*}{10} \right )^{-1/2} \left (\frac{\sigma}{10^9 {\rm GeV^3}} \right )^{-1} \left (\frac{c_1}{10^{-12} {\rm GeV^2}} \right )}} \quad {\rm GeV}
\end{align}
\end{widetext}
where $C_{\rm ann}\simeq 2$ is a dimensionless constant, $g_*$ is the relativistic degrees of freedom and $\mathcal{A}\simeq 0.8$ is the area parameter \cite{Hiramatsu:2013qaa}. If the bias term is absent or very small, domain walls will dominate the energy density of the Universe at a temperature given by \cite{Saikawa:2017hiv}
\begin{align}
    T_{\rm dom} &= 2.86\times10^{-6} \left (\frac{g_*}{10} \right)^{-1/4} \left (\frac{\mathcal{A}}{0.8} \right )^{1/2} \nonumber \\
    & \times \left (\frac{\sigma}{10^9\, {\rm GeV^3}} \right )^{1/2} {\rm GeV}.
\end{align}
Demanding the annihilation to occur before domination $T_{\rm ann} > T_{\rm dom}$, therefore, puts a lower bound on the bias term. We can impose another lower bound on the bias term by demanding the domain walls to annihilate before the BBN epoch $T_{\rm ann} > T_{\rm BBN}$ such that the light nuclei abundance does not get affected. On the other hand, the bias term can not be arbitrarily large as it would otherwise prevent the percolation of both the vacua separated by DW. This leads to an upper bound on the bias term $V_{\rm bias} < 0.795 V_0$ \cite{Saikawa:2017hiv} with $V_0$ being the height of the potential barrier separating the two minima.

The annihilation of DW can lead to the production of stochastic gravitational waves \cite{Vilenkin:1981zs, Gelmini:1988sf, Larsson:1996sp, Hiramatsu:2013qaa, Hiramatsu:2012sc, Saikawa:2017hiv,Roshan:2024qnv,Bhattacharya:2023kws}.  Earlier studies assumed that the DW annihilate instantaneously at $T=T_{\text{ann}}$ during radiation-dominated era. However, recent simulation studies~\cite{Kitajima:2023cek,Notari:2025kqq} indicate that the maximum production of gravitational waves occurs at a temperature lower than the annihilation temperature, namely
$T_{\rm gw} \simeq 0.3\, T_{\rm ann}\,$.
The present-day peak frequency $f_{p}$ and the corresponding peak amplitude $\Omega_{p} h^2$ of the gravitational wave signal have been extensively studied in Refs.~\cite{Kadota:2015dza, Hiramatsu:2013qaa, Saikawa:2017hiv, Chen:2020wvu}. Following the recent analyses in~\cite{Dankovsky:2024zvs,Notari:2025kqq,Blasi:2025tmn}, these quantities can be expressed as
\begin{widetext}
\begin{align} \label{fpeak}
f_{p}& \simeq 7.5\times 10^{-9} ~{\rm Hz}~\bigg(\frac{T_{\rm gw}}{0.1~\rm GeV} \bigg) \bigg(\frac{g_*(T_{\rm gw})}{10}\bigg)^{1/6},\nonumber \\
\Omega_{p}h^2 & \simeq 1\times 10^{-10} ~\bigg(\frac{0.1~ \rm GeV}{T_{\rm gw}}\bigg)^4 \bigg(\frac{\sigma}{10^{15}~{\rm GeV}^3}\bigg)^2 \bigg(\frac{10}{g_*(T_{\rm gw})}\bigg)^{4/3}.
\end{align}
% \begin{align}\label{fpeak}
%     f_{p}&\simeq 3.75\times10^{-9}~\text{Hz}\times C_{\rm ann}^{-1/2}\mathcal{A}^{-1/2}\bigg(\frac{10^3~\text{GeV}}{\mathcal{\sigma}^{1/3}}\bigg)^{3/2}\bigg(\frac{V_{\text{bias}}^{1/4}}{10^{-3}~\text{GeV}}\bigg)^{2}\,,\nonumber\\
%     \Omega_{p}h^2&\simeq 5.3\times10^{-20}\times\tilde{\epsilon}_{\text{GW}}~C_{\rm ann}^{2}\mathcal{A}^{4}\bigg(\frac{\sigma^{1/3}}{10^3~\text{GeV}}\bigg)^{12}\bigg(\frac{10^{-3}~\text{GeV}}{V_{\text{bias}}^{1/4}}\bigg)^{8},
% \end{align}
 \end{widetext}
 %where $\tilde{\epsilon}_{\text{GW}}\simeq0.7$~\cite{Hiramatsu:2013qaa} is the fraction of energy radiated into GW.}
 The GW spectrum from DW follows a broken power law, where the
breaking point has a frequency determined by the annihilation time and the peak amplitude
is determined by the energy density in the domain walls as can be seen from Eq. \eqref{fpeak}. The broken power-law spectrum can be parametrised as~\cite{Caprini:2019egz, NANOGrav:2023hvm}
\begin{eqnarray}
 \Omega_{\rm GW}h^2_{} = \Omega_p h^2 \frac{(a+b)^c}{\left(b x^{-a / c}+a x^{b / c}\right)^c} \ ,
\label{eq:spec-par}
\end{eqnarray}
where $x \equiv f/f_p$, and $a$, $b$ and $c$ are real and positive parameters. Here the low-frequency slope $a = 3$ can be fixed by causality, while numerical simulations suggest $b \simeq c \simeq 1$~\cite{Hiramatsu:2013qaa}. The bounds on effective relativistic degrees of freedom $N_{\rm eff}$ from CMB and BBN observations lead to an upper limit on maximum GW amplitude as $ \Omega_{\rm GW}^{\rm \Delta N_{\rm eff}}h^2 \lesssim 1.75\times 10^{-6}$ \cite{Domenech:2020ssp, Planck:2018vyg}.

\begin{figure*}
    \centering
    \includegraphics[scale=0.5]{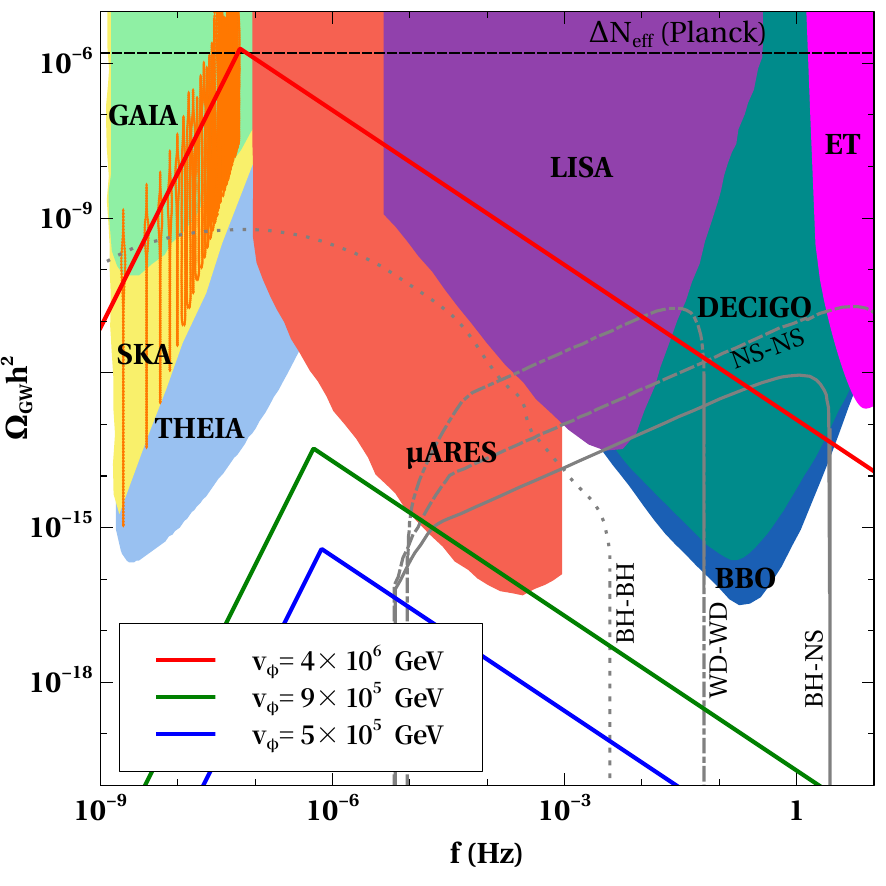}
        \includegraphics[scale=0.5]{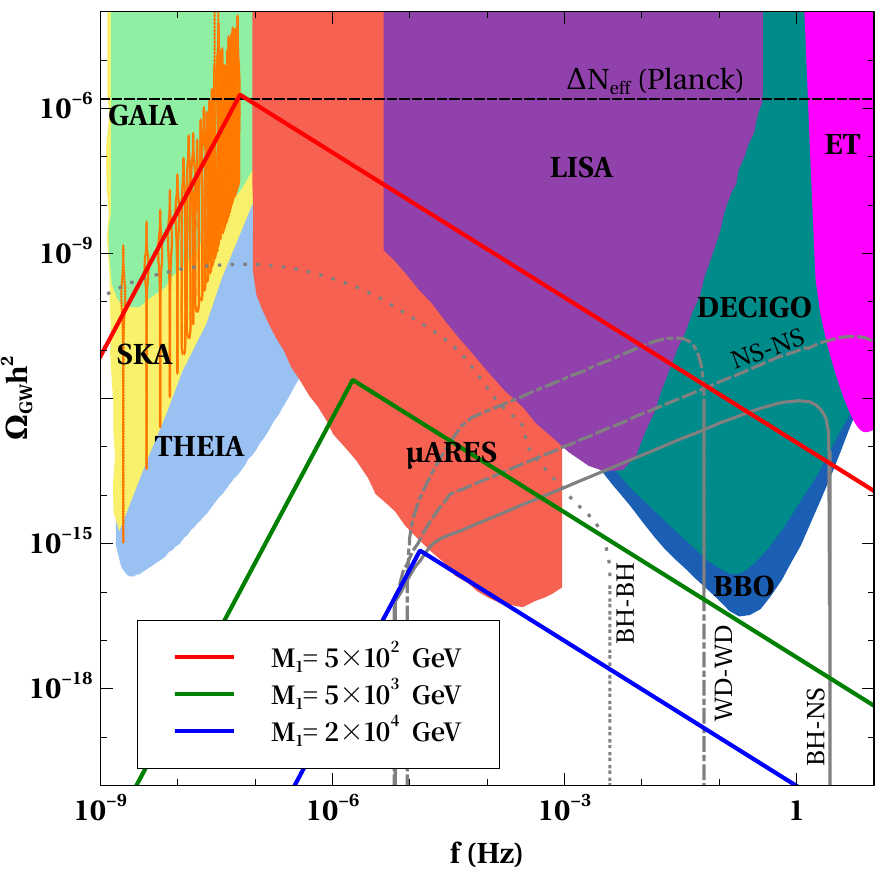} \\
            \includegraphics[scale=0.5]{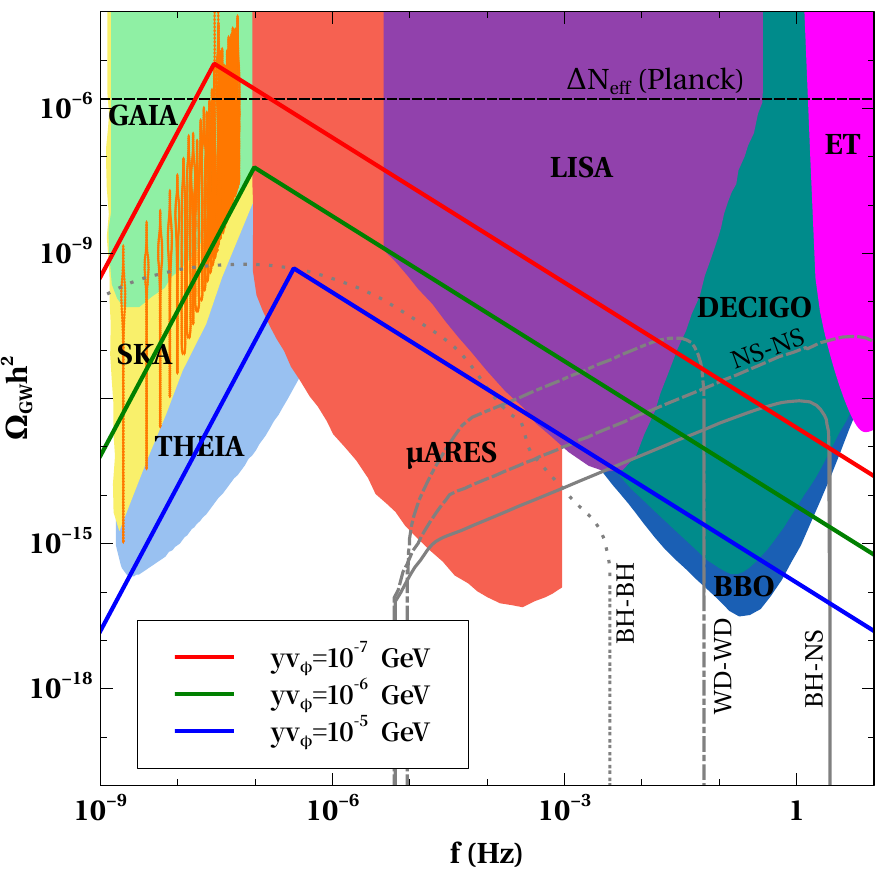}
             \includegraphics[scale=0.5]{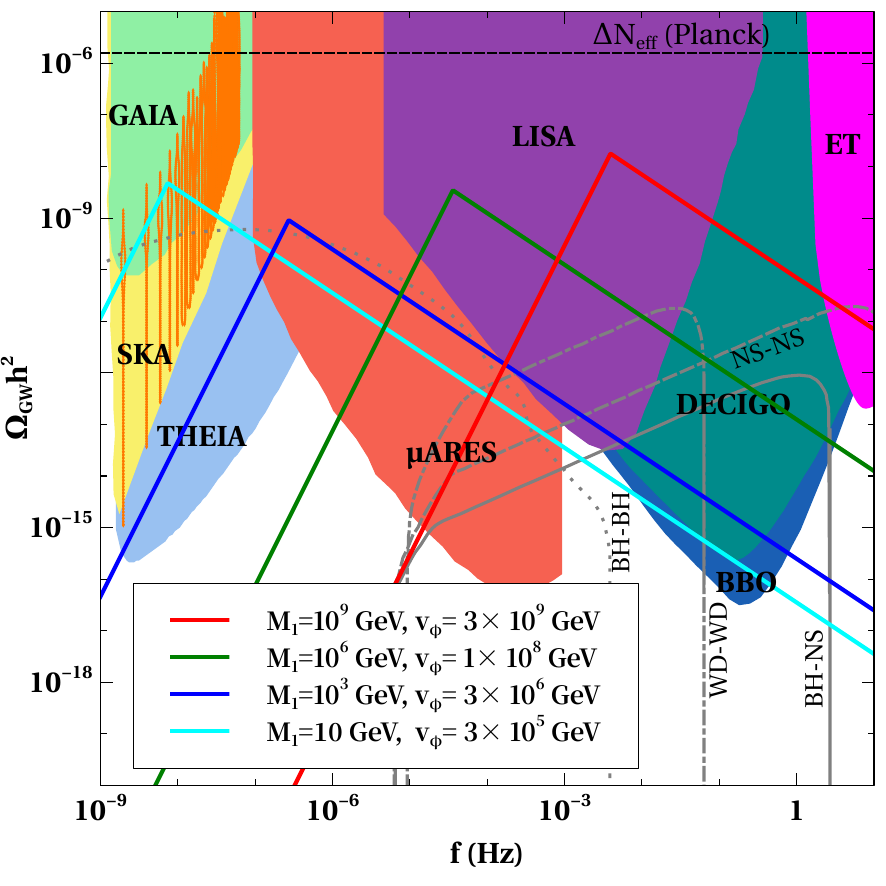}
    \caption{GW spectrum with $M_1$=500 GeV and $y v_\phi\sim 1$ keV (top left), $v_\phi=4 \times 10^6$ GeV and  $yv_\phi=1$ keV (top right), $M_1$=500 GeV and $v_\phi=3 \times 10^6$ GeV (bottom left), different combinations of $M_1, v_\phi$ and fixed $y v_\phi\sim 1$ keV (bottom right). The shaded regions correspond to future GW experiments' sensitivities with orange colored data points correspond to NANOGrav 2023 data. The horizontal black dashed line corresponds to the upper limit on GW contribution to dark radiation or $\Delta N_{\rm eff}$. The grey colored contours correspond to different astrophysical foregrounds. See text for details.}
    \label{fig:gw1}
\end{figure*}

\begin{figure*}
    \centering
   \includegraphics[scale=0.36]{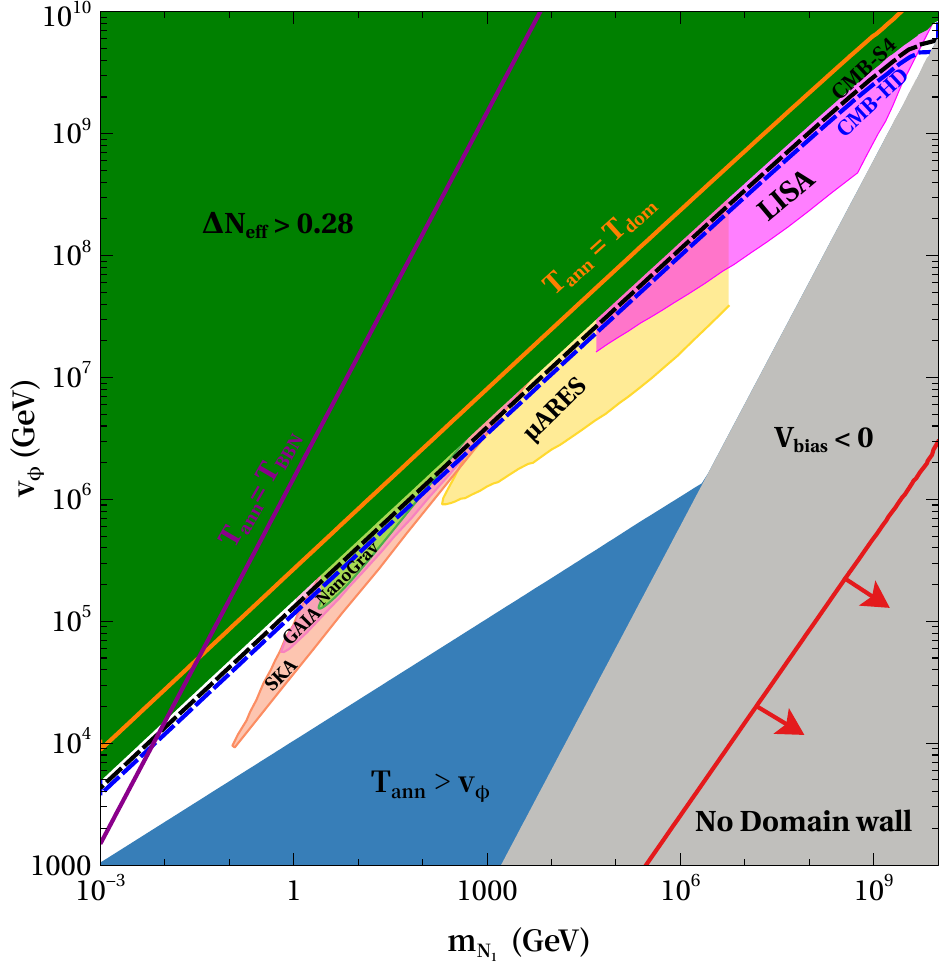}
      \includegraphics[scale=0.39]{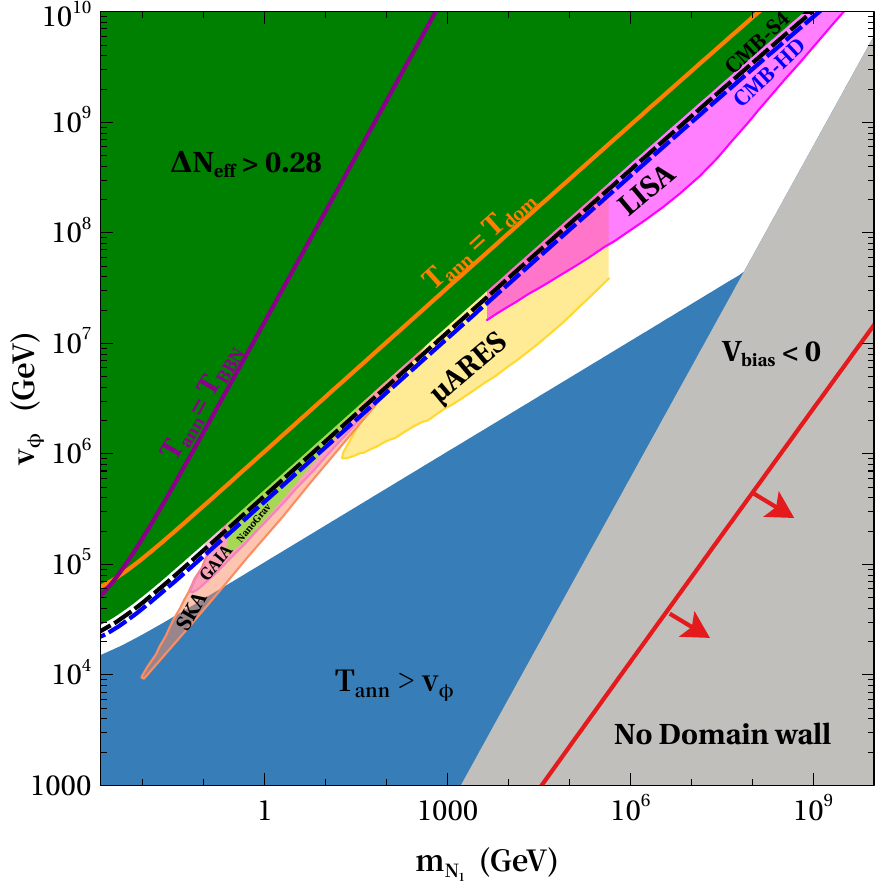}
      \includegraphics[scale=0.39]{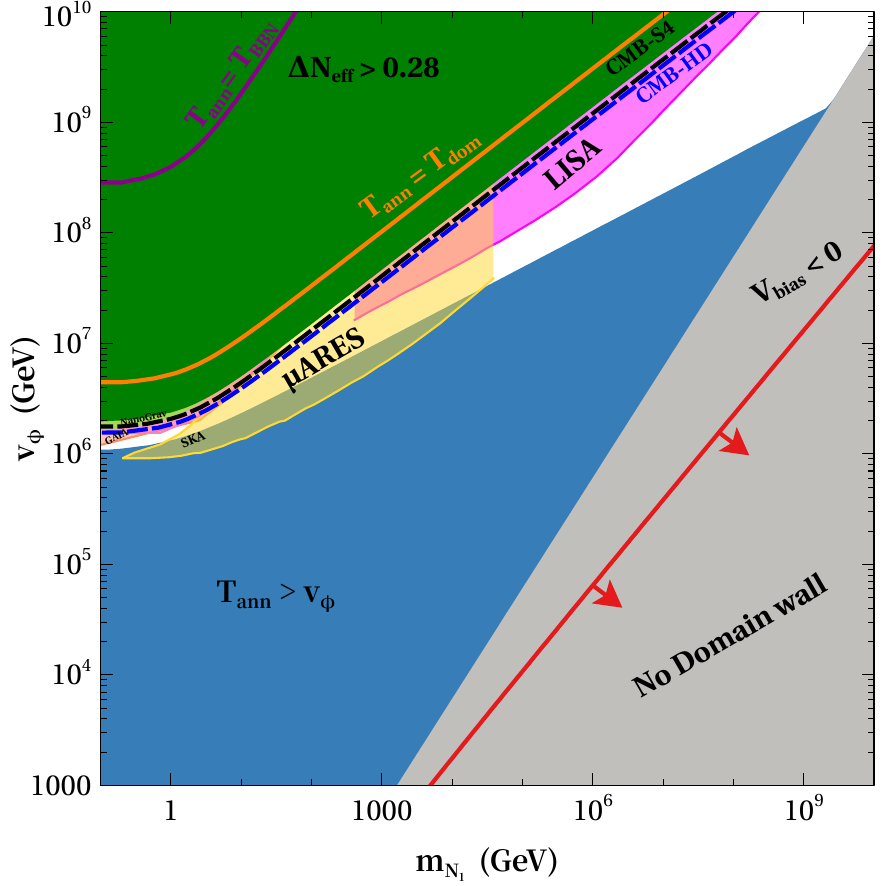}
    \caption{Parameter space in $m_{N_1}$-$v_\phi$ plane, with $y v_\phi =1$ keV (left), $y v_\phi =1$ MeV (middle) and $y v_\phi =1$ GeV (right). The white region towards the right of the solid purple contour remains allowed part of which can be probed at future CMB and GW experiments.}
    \label{fig:1}
\end{figure*}

Fig. \ref{fig:gw1} shows the GW spectra for different model parameters. Depending upon the scale of $Z_2$-breaking, scale of seesaw and $\phi-N_i$ couplings, the peak amplitude and peak frequency of the spectrum can lie within reach of several present and future experiments. Shaded colored regions show the sensitivities of various future GW experiments like BBO~\cite{Yagi:2011wg}, LISA~\cite{2017arXiv170200786A}, ET~\cite{Punturo_2010}, THEIA~\cite{Garcia-Bellido:2021zgu}, $\mu$ARES~\cite{Sesana:2019vho}, SKA~\cite{Weltman:2018zrl}, GAIA \cite{Garcia-Bellido:2021zgu} while the orange colored violin-shaped points show the NANOGrav (NG) results \cite{NANOGrav:2023gor}. While various astrophysical sources including compact binaries such as supermassive black hole binaries (BH–BH), white dwarf binaries (WD-WD), black hole–neutron star systems (BH-NS), and neutron star–neutron star binaries (NS-NS) are also expected to generate gravitational wave signal, the precise estimation of their contributions requires a detailed analysis of the most probable mass distributions and coalescence rates \cite{Rosado:2011kv}, which we do not attempt here. Supermassive black hole binaries predominantly generate gravitational waves in the low-frequency regime, making them a plausible explanation for the recently reported signal by the NANOGrav collaboration~\cite{NANOGrav:2023gor}. Nevertheless, statistical analysis indicates that a combination of astrophysical and cosmological sources may provide a better fit to the observed data \cite{NANOGrav:2023hvm}. In contrast, other compact binary populations typically produce foregrounds at higher frequencies, potentially detectable by LISA, DECIGO, BBO, $\mu$ARES, and ET. We show these astrophysical foregrounds as grey colored contours in Fig. \ref{fig:gw1}. However, only detailed post-detection analyses will allow a definitive determination of the underlying sources, which is beyond the scope of the present work. We leave a detailed analysis of the gravitational wave signal based on the signal-to-noise ratio, including the impact of astrophysical foregrounds, following the discussion of Ref.~\cite{Baldes:2023rqv}, for future work. Fig. \ref{fig:1} shows the parameter space in the plane of seesaw scale $M_1$ and $Z_2$-breaking scale $v_\phi$ for three different values of $y v_\phi=1$ keV (left), $y v_\phi =1$ MeV (middle) and $y v_\phi =1$ GeV (right). The green shaded region is disfavored from PLANCK 2018 bound $N_{\rm eff}=2.99^{+0.34}_{-0.33}$ at $2\sigma$ CL \cite{Planck:2018vyg} including baryon acoustic oscillation (BAO) data. The blue shaded region is disfavored from the criteria that DW should annihilate at a temperature below the symmetry breaking scale. The grey shaded region is disfavored due to our choice of true minima. The solid red line indicates the upper bound on the bias term from percolation such that for the parameter space below this line, domain walls never form. The solid purple and solid orange contours correspond to $T_{\rm ann}=T_{\rm BBN}, T_{\rm ann}=T_{\rm dom}$ respectively, such that the regions towards the left of these contours are disfavored. The dashed contours correspond to future sensitivity of CMB Stage IV (CMB-S4): $\Delta {\rm N}_{\rm eff}={\rm N}_{\rm eff}-{\rm N}^{\rm SM}_{\rm eff}
= 0.06$ \cite{Abazajian:2019eic} and CMB-HD \cite{CMB-HD:2022bsz}: $\Delta N_{\rm eff}$ upto $0.014$ at $1\sigma$. The remaining shaded regions correspond to sensitivities of present and future GW experiments like NanoGrav \cite{NANOGrav:2023gor}, LISA~\cite{2017arXiv170200786A}, $\mu$ARES~\cite{Sesana:2019vho}, SKA~\cite{Weltman:2018zrl} and GAIA \cite{Garcia-Bellido:2021zgu} to the peak frequencies and peak amplitudes of the GW spectrum. As $y v_\phi$ increases, the bias also increases decreasing the peak GW amplitude and hence weakening the $N_{\rm eff}$ bounds. However, the blue region expands with larger bias as $T_{\rm ann}$ increases with increase in bias. For the right panel plot with $y v_\phi=1$ GeV, the $N_{\rm eff}$ as well as $T=T_{\rm BBN}, T=T_{\rm dom}$ contours show weaker dependence on $M_1$ in low mass region as the mass of RHN starts getting dominated by the $y v_\phi$ term. This explains the behavior of the three plots shown in Fig. \ref{fig:1} for three different values of $y v_\phi$. To summarize, a large part of the currently allowed parameter space can be probed at CMB and GW experiments.

\begin{figure}
    \centering
    \includegraphics[scale=0.5]{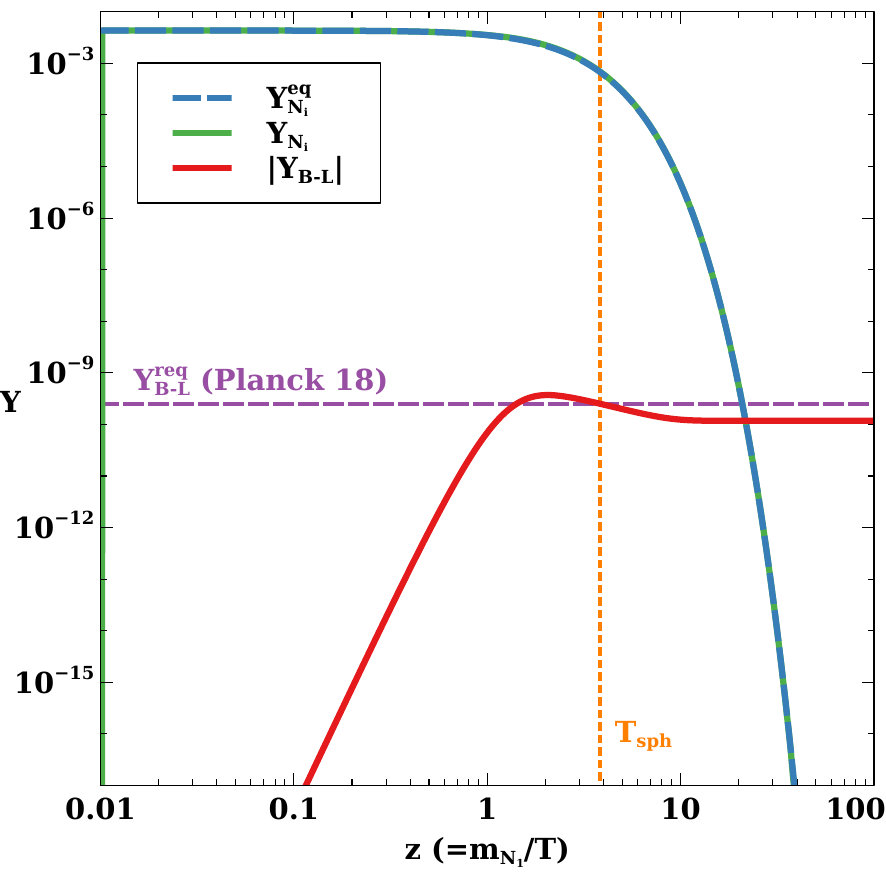}
    \caption{Evolution of comoving number densities for $m_{N_1} = 500$ GeV and $m_{N_1}-m_{N_2}= y v_\phi =1$ keV. The complex angle in CI parametrisation is chosen as $z_1=0.28+0.15i$.}
    \label{fig:evo}
\end{figure}

\section{Leptogenesis}
\label{sec3}
We briefly outline the possibility of low scale leptogenesis in our scenario assuming the small mass splitting among RHNs to arise from their couplings to the $Z_2$-odd scalar $\phi$. The CP asymmetry parameter for $N_i$ decay (summing over all lepton flavours) is given by \cite{Pilaftsis:2003gt}
\begin{equation}\begin{aligned}
\epsilon_{i} & = \dfrac{\Gamma(N_{i}\to\sum_{\alpha}L_{\alpha} H )-\Gamma(N_{i}\to \sum_{\alpha}\bar L_{\alpha}\bar{H})}{\Gamma(N_{i}\to \sum_{\alpha} L_{\alpha}H)+\Gamma(N_{i}\to\sum_{i}\bar L_{\alpha}\bar{H})}  \\ 
 & = \dfrac{{\rm Im}[(h^{\dagger} h)_{ij}^{2}]}{(h^{\dagger} h)_{ii}(h^{\dagger} h)_{jj}}\dfrac{(m_{N_i}^{2}-m_{N_j}^{2})m_{N_i}\Gamma_{j}}{(m_{N_i}^{2}-m_{N_j}^{2})^{2}+m_{N_i}^{2}\Gamma_{j}^{2}}. \label{eq:asymmparameter}
 \end{aligned}\end{equation}
Since we are mainly focusing on the parameter space where $m_{N_1}\simeq m_{N_2}$ such that the leptogenesis is mainly governed by the resonant enhancement between $N_{1}$ and $N_{2}$ provided $m_{N_2}-m_{N_1} \simeq \Gamma_1/2$ where $\Gamma_1$ is the decay width of the lightest RHN $N_1$.

The relevant Boltzmann equations for our setup can be written as 
\begin{equation}
\begin{aligned}
\dfrac{dY_{N_{i}}}{dz} & = D_{i}\Delta Y_{N_i} - \dfrac{s}{{\bf H}z}\Delta Y^2_{N_i}\langle \sigma v\rangle_{\small{N_{i}N_{i}\to X X}}\\ 
&- \dfrac{s}{{\bf H}z}\Delta Y_{N_i}Y^{\rm eq}_{SM}\langle \sigma v\rangle_{\small{N_{i}SM\to SM \,SM}},  \\
\dfrac{dY_{B-L}}{dz} &= -\sum_{i}\epsilon_{i}D_{i}\Delta Y_{N_i}-\left(\Delta W+\sum_{i}W_{i}\right)Y_{B-L}, \label{eq:B-L}
\end{aligned}
\end{equation}
where
\begin{equation}
\Delta Y_{N_i}=Y_{N_i}-Y_{N_i}^{\rm eq},~
\Delta Y^2_{N_i}=Y^2_{N_i}-(Y_{N_i}^{\rm eq})^2.
\end{equation}
$Y_{N_i}(Y_{B-L})$ denote the comoving number densities of $N_{i}(B-L)$, with $i=1,2$, $z=m_{N_1}/T=m_{N_2}/T$. The entropy density of the Universe is denoted by $s=\frac{2\pi^2}{45}g_{*s}T^3$ with $g_{*s}$ being the relativistic entropy degrees of freedom. In $\langle \sigma v \rangle_{N_i N_i \rightarrow X X}$, $X$ denotes any final state particle to which $N_i$'s can annihilate into. Similarly, in $\langle \sigma v\rangle_{\small{N_{i}SM\to SM \,SM}}$, we consider the scatterings with one $N_i$ as external state which be dominant at higher temperatures, bringing $N_i$ into equilibrium quickly \cite{Giudice:2003jh, Hahn-Woernle:2009jyb}. $D_{1,2}$ are the decay terms and $W_{i}$ is the inverse decay washout term for $N_{i}$, defined as
\begin{eqnarray}
D_{i} = \dfrac{ \langle \Gamma_{i} \rangle}{{\bf H} z} = K_{i}z\dfrac{\kappa_{1}(z)}{\kappa_{2}(z)}, \,\,
W_{i} = \dfrac{1}{4}K_{i}z^{3}\kappa_{1}(z)
\end{eqnarray}
with $K_i=\Gamma_i/{\bf H}(z=1)$ being the decay parameter and $\kappa_i(z)$ being the modified Bessel function of $i$-th kind. The term $\Delta W$ on the right-hand side of the Eq. \eqref{eq:B-L} takes account of all the scattering processes that can act as possible washouts for the generated $B-L$ asymmetry. Fig. \ref{fig:evo} shows the evolution of comoving number densities of $N_1$ and $B-L$ as a function of $z$ considering $m_{N_1} = 500$ GeV and $m_{N_1}-m_{N_2}= y v_\phi =1$ keV. The vertical dashed line corresponds to the sphaleron decoupling epoch $T_{\rm sph} \sim 130$ GeV. The lepton asymmetry generated till the sphaleron decoupling epoch gets converted into baryon asymmetry as~\cite{Harvey:1990qw}
\begin{align}
Y_B (T_{\rm sph}) \simeq a_\text{Sph}\,Y_{B-L} (T_{\rm sph})
\label{eq:sphaleron}
\end{align}
where $a_\text{Sph}=\frac{8\,N_F+4\,N_H}{22\,N_F+13\,N_H} = \frac{28}{79} Y_{B-L}$ is the sphaleron conversion factor with $N_F=3\,,N_H=1$ being the fermion generations and the number of scalar doublets respectively. The observational constraint on baryon-to-photon ratio $\eta_B =(6.12\pm 0.04) \times 10^{-10}$ \cite{Planck:2018vyg} leads to the following requirement on $B-L$ asymmetry 
\begin{equation}
    Y_{B-L} (T_{\rm sph}) =  (2.45\pm 0.02) \times 10^{-10},
    \label{eq:3.4}
\end{equation}
which is shown by the horizontal dashed line in Fig. \ref{fig:evo}. Clearly, the benchmark point chosen here is consistent with the observed baryon asymmetry. In addition to the type-I seesaw related interactions incorporated in the above Boltzmann equations, the presence of $\Phi$ can introduce additional production and annihilation channels of RHNs. In particular, $\Phi \rightarrow N_i N_i$ can produce $N_i$'s and $N_i N_i \rightarrow H H$ mediated by $\Phi$ can lead to annihilations of $N_i$. We have checked that for the benchmark points discussed here, the additional production from $\Phi$ does not change the numerical results shown in Fig. \ref{fig:evo}. On the other hand, the annihilation into a pair of Higgs can be kept suppressed by tuning the $\Phi-H$ quartic coupling, a free parameter which does not affect rest of our analysis.

It should be noted that for the chosen benchmark point where the strength of the bias also gives keV scale mass splittings of RHNs required for resonant leptogenesis, the scale of leptogenesis precedes the annihilation of DW. As a representative value, the annihilation temperature for this benchmark is found to be $T_{\rm ann}=121.87~{\rm GeV}$ by fixing the symmetry breaking scale at $v_\phi=10^5~{\rm GeV}$. On the other hand, DW never dominates the energy density of the Universe in our scenario. Therefore, the annihilation of DW at a later stage does not change the asymmetry generated at a higher scale. DW dominance before annihilation \cite{Kawasaki:2004rx, Hattori:2015xla, Hong:2025piv} can change the conclusions of our scenario which we leave for future studies.

\section{Conclusion}
\label{sec4}
We have studied the possibility of $Z_2$ domain walls getting destabilized by interactions involving heavy right-handed neutrinos which are also responsible for generating light neutrino masses via the seesaw mechanism. A $Z_2$-odd real scalar singlet $\phi$ acquires a non-zero VEV leading to spontaneous $Z_2$ breaking and hence the formation of domain walls. We then introduce explicit $Z_2$ symmetry breaking interactions of $\phi$ with heavy RHN $N_i$. While these RHNs also lead to type-I seesaw origin of light neutrino masses via coupling to lepton doublets, they also generate $Z_2$-breaking terms in the scalar potential of $\phi$ via radiative corrections. We show that depending upon the seesaw scale and the strength of $\phi$-RHN coupling, the bias term required for domain wall annihilation can be generated. The resulting gravitational wave spectrum from domain wall annihilation can remain within reach of several present and future experiments. The seesaw link to the bias term also offers unique correlation between GW spectrum and the seesaw scale, leading to an indirect detection probe of the latter. Since $\phi$-RHN coupling remains typically small for the required bias, it can also give rise to small mass splitting among RHNs assuming they remain degenerate at leading order. While our minimal setup does not have this restriction, in UV complete models with additional flavor symmetry \cite{Borah:2017qdu} it is possible to have degenerate RHN masses at leading order with additional contributions arising at higher orders responsible for tiny mass splittings. We show that such tiny mass splittings can naturally lead to resonant leptogenesis, consistent with the observed baryon asymmetry of the Universe. Another possible direction is to look for UV completions where explicit $Z_2$-breaking terms with odd powers of $\phi$ are disallowed in the tree level scalar potential, motivating the role of radiative generation. This is similar to some of the proposed solutions of the axion quality problem with additional gauge symmetries which prevent certain global symmetry breaking operators \cite{Barr:1992qq}. We leave the exploration of such UV complete scenarios to future studies.

\acknowledgments
The work of D.B. is supported by the Science and Engineering Research Board (SERB), Government of India grants MTR/2022/000575, CRG/2022/000603. D.B. also thanks Subhojit Roy for useful discussions on collapsing domain walls. I.S. acknowledges the support from SERB, Government of India grant CRG/2022/000603.

\bibliographystyle{JHEP}
%\bibstyle{apsrev}
\bibliography{arxiv_final}

\providecommand{\href}[2]{#2}\begingroup\raggedright\begin{thebibliography}{10}

\bibitem{ParticleDataGroup:2024cfk}
{\bf Particle Data Group} Collaboration, S.~Navas {\em et~al.}, {\it {Review of particle physics}},  {\em Phys. Rev. D} {\bf 110} (2024), no.~3 030001.

\bibitem{Minkowski:1977sc}
P.~Minkowski, {\it {$\mu \to e\gamma$ at a Rate of One Out of $10^{9}$ Muon Decays?}},  {\em Phys. Lett. B} {\bf 67} (1977) 421--428.

\bibitem{GellMann:1980vs}
M.~Gell-Mann, P.~Ramond, and R.~Slansky, {\it {Complex Spinors and Unified Theories}},  {\em Conf. Proc. C} {\bf 790927} (1979) 315--321, [\href{http://arxiv.org/abs/1306.4669}{{\tt arXiv:1306.4669}}].

\bibitem{Mohapatra:1979ia}
R.~N. Mohapatra and G.~Senjanovic, {\it {Neutrino Mass and Spontaneous Parity Nonconservation}},  {\em Phys. Rev. Lett.} {\bf 44} (1980) 912.

\bibitem{Sawada:1979dis}
O.~Sawada and A.~Sugamoto, eds., {\em {Proceedings: Workshop on the Unified Theories and the Baryon Number in the Universe}: {Tsukuba, Japan, February 13-14, 1979}}, (Tsukuba, Japan), Natl.Lab.High Energy Phys., 1979.

\bibitem{Yanagida:1980xy}
T.~Yanagida, {\it {Horizontal Symmetry and Masses of Neutrinos}},  {\em Prog. Theor. Phys.} {\bf 64} (1980) 1103.

\bibitem{Schechter:1980gr}
J.~Schechter and J.~Valle, {\it {Neutrino Masses in SU(2) x U(1) Theories}},  {\em Phys. Rev. D} {\bf 22} (1980) 2227.

\bibitem{Planck:2018vyg}
{\bf Planck} Collaboration, N.~Aghanim {\em et~al.}, {\it {Planck 2018 results. VI. Cosmological parameters}},  {\em Astron. Astrophys.} {\bf 641} (2020) A6, [\href{http://arxiv.org/abs/1807.06209}{{\tt arXiv:1807.06209}}]. [Erratum: Astron.Astrophys. 652, C4 (2021)].

\bibitem{Fukugita:1986hr}
M.~Fukugita and T.~Yanagida, {\it {Baryogenesis Without Grand Unification}},  {\em Phys. Lett. B} {\bf 174} (1986) 45--47.

\bibitem{Sakharov:1967dj}
A.~D. Sakharov, {\it {Violation of CP Invariance, C asymmetry, and baryon asymmetry of the universe}},  {\em Pisma Zh. Eksp. Teor. Fiz.} {\bf 5} (1967) 32--35. [Usp. Fiz. Nauk161,no.5,61(1991)].

\bibitem{Kuzmin:1985mm}
V.~A. Kuzmin, V.~A. Rubakov, and M.~E. Shaposhnikov, {\it {On the Anomalous Electroweak Baryon Number Nonconservation in the Early Universe}},  {\em Phys. Lett.} {\bf 155B} (1985) 36.

\bibitem{Davidson:2002qv}
S.~Davidson and A.~Ibarra, {\it {A Lower bound on the right-handed neutrino mass from leptogenesis}},  {\em Phys. Lett.} {\bf B535} (2002) 25--32, [\href{http://arxiv.org/abs/hep-ph/0202239}{{\tt hep-ph/0202239}}].

\bibitem{Pilaftsis:2003gt}
A.~Pilaftsis and T.~E.~J. Underwood, {\it {Resonant leptogenesis}},  {\em Nucl. Phys. B} {\bf 692} (2004) 303--345, [\href{http://arxiv.org/abs/hep-ph/0309342}{{\tt hep-ph/0309342}}].

\bibitem{Shaposhnikov:2006nn}
M.~Shaposhnikov, {\it {A Possible symmetry of the nuMSM}},  {\em Nucl. Phys. B} {\bf 763} (2007) 49--59, [\href{http://arxiv.org/abs/hep-ph/0605047}{{\tt hep-ph/0605047}}].

\bibitem{Kersten:2007vk}
J.~Kersten and A.~Y. Smirnov, {\it {Right-Handed Neutrinos at CERN LHC and the Mechanism of Neutrino Mass Generation}},  {\em Phys. Rev. D} {\bf 76} (2007) 073005, [\href{http://arxiv.org/abs/0705.3221}{{\tt arXiv:0705.3221}}].

\bibitem{Moffat:2017feq}
K.~Moffat, S.~Pascoli, and C.~Weiland, {\it {Equivalence between massless neutrinos and lepton number conservation in fermionic singlet extensions of the Standard Model}},  \href{http://arxiv.org/abs/1712.07611}{{\tt arXiv:1712.07611}}.

\bibitem{Klaric:2020phc}
J.~Klari{\'c}, M.~Shaposhnikov, and I.~Timiryasov, {\it {Uniting Low-Scale Leptogenesis Mechanisms}},  {\em Phys. Rev. Lett.} {\bf 127} (2021), no.~11 111802, [\href{http://arxiv.org/abs/2008.13771}{{\tt arXiv:2008.13771}}].

\bibitem{Drewes:2021nqr}
M.~Drewes, Y.~Georis, and J.~Klari\'c, {\it {Mapping the Viable Parameter Space for Testable Leptogenesis}},  {\em Phys. Rev. Lett.} {\bf 128} (2022), no.~5 051801, [\href{http://arxiv.org/abs/2106.16226}{{\tt arXiv:2106.16226}}].

\bibitem{Hernandez:2022ivz}
P.~Hernandez, J.~Lopez-Pavon, N.~Rius, and S.~Sandner, {\it {Bounds on right-handed neutrino parameters from observable leptogenesis}},  {\em JHEP} {\bf 12} (2022) 012, [\href{http://arxiv.org/abs/2207.01651}{{\tt arXiv:2207.01651}}].

\bibitem{Dror:2019syi}
J.~A. Dror, T.~Hiramatsu, K.~Kohri, H.~Murayama, and G.~White, {\it {Testing the Seesaw Mechanism and Leptogenesis with Gravitational Waves}},  {\em Phys. Rev. Lett.} {\bf 124} (2020), no.~4 041804, [\href{http://arxiv.org/abs/1908.03227}{{\tt arXiv:1908.03227}}].

\bibitem{Blasi:2020wpy}
S.~Blasi, V.~Brdar, and K.~Schmitz, {\it {Fingerprint of low-scale leptogenesis in the primordial gravitational-wave spectrum}},  {\em Phys. Rev. Res.} {\bf 2} (2020), no.~4 043321, [\href{http://arxiv.org/abs/2004.02889}{{\tt arXiv:2004.02889}}].

\bibitem{Fornal:2020esl}
B.~Fornal and B.~Shams Es~Haghi, {\it {Baryon and Lepton Number Violation from Gravitational Waves}},  {\em Phys. Rev. D} {\bf 102} (2020), no.~11 115037, [\href{http://arxiv.org/abs/2008.05111}{{\tt arXiv:2008.05111}}].

\bibitem{Samanta:2020cdk}
R.~Samanta and S.~Datta, {\it {Gravitational wave complementarity and impact of NANOGrav data on gravitational leptogenesis}},  {\em JHEP} {\bf 05} (2021) 211, [\href{http://arxiv.org/abs/2009.13452}{{\tt arXiv:2009.13452}}].

\bibitem{Barman:2022yos}
B.~Barman, D.~Borah, A.~Dasgupta, and A.~Ghoshal, {\it {Probing High Scale Dirac Leptogenesis via Gravitational Waves from Domain Walls}},  \href{http://arxiv.org/abs/2205.03422}{{\tt arXiv:2205.03422}}.

\bibitem{Huang:2022vkf}
P.~Huang and K.-P. Xie, {\it {Leptogenesis triggered by a first-order phase transition}},  \href{http://arxiv.org/abs/2206.04691}{{\tt arXiv:2206.04691}}.

\bibitem{Dasgupta:2022isg}
A.~Dasgupta, P.~S.~B. Dev, A.~Ghoshal, and A.~Mazumdar, {\it {Gravitational Wave Pathway to Testable Leptogenesis}},  \href{http://arxiv.org/abs/2206.07032}{{\tt arXiv:2206.07032}}.

\bibitem{Okada:2018xdh}
N.~Okada and O.~Seto, {\it {Probing the seesaw scale with gravitational waves}},  {\em Phys. Rev. D} {\bf 98} (2018), no.~6 063532, [\href{http://arxiv.org/abs/1807.00336}{{\tt arXiv:1807.00336}}].

\bibitem{Hasegawa:2019amx}
T.~Hasegawa, N.~Okada, and O.~Seto, {\it {Gravitational waves from the minimal gauged $U(1)_{B-L}$ model}},  {\em Phys. Rev. D} {\bf 99} (2019), no.~9 095039, [\href{http://arxiv.org/abs/1904.03020}{{\tt arXiv:1904.03020}}].

\bibitem{Borah:2022cdx}
D.~Borah, A.~Dasgupta, and I.~Saha, {\it {Leptogenesis and dark matter through relativistic bubble walls with observable gravitational waves}},  {\em JHEP} {\bf 11} (2022) 136, [\href{http://arxiv.org/abs/2207.14226}{{\tt arXiv:2207.14226}}].

\bibitem{Borah:2022vsu}
D.~Borah, S.~Jyoti~Das, and R.~Roshan, {\it {Probing high scale seesaw and PBH generated dark matter via gravitational waves with multiple tilts}},  \href{http://arxiv.org/abs/2208.04965}{{\tt arXiv:2208.04965}}.

\bibitem{Barman:2023fad}
B.~Barman, D.~Borah, S.~Jyoti~Das, and I.~Saha, {\it {Scale of Dirac leptogenesis and left-right symmetry in the light of recent PTA results}},  {\em JCAP} {\bf 10} (2023) 053, [\href{http://arxiv.org/abs/2307.00656}{{\tt arXiv:2307.00656}}].

\bibitem{Borah:2023saq}
D.~Borah, A.~Dasgupta, and I.~Saha, {\it {LIGO-VIRGO constraints on dark matter and leptogenesis triggered by a first order phase transition at high scale}},  \href{http://arxiv.org/abs/2304.08888}{{\tt arXiv:2304.08888}}.

\bibitem{Saikawa:2017hiv}
K.~Saikawa, {\it {A review of gravitational waves from cosmic domain walls}},  {\em Universe} {\bf 3} (2017), no.~2 40, [\href{http://arxiv.org/abs/1703.02576}{{\tt arXiv:1703.02576}}].

\bibitem{Roshan:2024qnv}
R.~Roshan and G.~White, {\it {Using gravitational waves to see the first second of the Universe}},  \href{http://arxiv.org/abs/2401.04388}{{\tt arXiv:2401.04388}}.

\bibitem{Bhattacharya:2023kws}
S.~Bhattacharya, N.~Mondal, R.~Roshan, and D.~Vatsyayan, {\it {Leptogenesis, dark matter and gravitational waves from discrete symmetry breaking}},  {\em JCAP} {\bf 06} (2024) 029, [\href{http://arxiv.org/abs/2312.15053}{{\tt arXiv:2312.15053}}].

\bibitem{Blasi:2022ayo}
S.~Blasi, A.~Mariotti, A.~Rase, A.~Sevrin, and K.~Turbang, {\it {Friction on ALP domain walls and gravitational waves}},  {\em JCAP} {\bf 04} (2023) 008, [\href{http://arxiv.org/abs/2210.14246}{{\tt arXiv:2210.14246}}].

\bibitem{Blasi:2023sej}
S.~Blasi, A.~Mariotti, A.~Rase, and A.~Sevrin, {\it {Axionic domain walls at Pulsar Timing Arrays: QCD bias and particle friction}},  {\em JHEP} {\bf 11} (2023) 169, [\href{http://arxiv.org/abs/2306.17830}{{\tt arXiv:2306.17830}}].

\bibitem{Borah:2024kfn}
D.~Borah, N.~Das, and R.~Roshan, {\it {Observable gravitational waves and {\ensuremath{\Delta}}Neff with global lepton number symmetry and dark matter}},  {\em Phys. Rev. D} {\bf 110} (2024), no.~7 075042, [\href{http://arxiv.org/abs/2406.04404}{{\tt arXiv:2406.04404}}].

\bibitem{Borboruah:2024lli}
Z.~A. Borboruah, D.~Borah, L.~Malhotra, and U.~Patel, {\it {Minimal Dirac seesaw dark matter}},  {\em Phys. Rev. D} {\bf 112} (2025), no.~1 015022, [\href{http://arxiv.org/abs/2412.12267}{{\tt arXiv:2412.12267}}].

\bibitem{Zeldovich:1974uw}
Y.~B. Zeldovich, I.~Y. Kobzarev, and L.~B. Okun, {\it {Cosmological Consequences of the Spontaneous Breakdown of Discrete Symmetry}},  {\em Zh. Eksp. Teor. Fiz.} {\bf 67} (1974) 3--11.

\bibitem{Vilenkin:1981zs}
A.~Vilenkin, {\it {Gravitational Field of Vacuum Domain Walls and Strings}},  {\em Phys. Rev. D} {\bf 23} (1981) 852--857.

\bibitem{Sikivie:1982qv}
P.~Sikivie, {\it {Of Axions, Domain Walls and the Early Universe}},  {\em Phys. Rev. Lett.} {\bf 48} (1982) 1156--1159.

\bibitem{Gelmini:1988sf}
G.~B. Gelmini, M.~Gleiser, and E.~W. Kolb, {\it {Cosmology of Biased Discrete Symmetry Breaking}},  {\em Phys. Rev. D} {\bf 39} (1989) 1558.

\bibitem{Larsson:1996sp}
S.~E. Larsson, S.~Sarkar, and P.~L. White, {\it {Evading the cosmological domain wall problem}},  {\em Phys. Rev. D} {\bf 55} (1997) 5129--5135, [\href{http://arxiv.org/abs/hep-ph/9608319}{{\tt hep-ph/9608319}}].

\bibitem{Rai:1992xw}
B.~Rai and G.~Senjanovic, {\it {Gravity and domain wall problem}},  {\em Phys. Rev. D} {\bf 49} (1994) 2729--2733, [\href{http://arxiv.org/abs/hep-ph/9301240}{{\tt hep-ph/9301240}}].

\bibitem{Lew:1993yt}
H.~Lew and A.~Riotto, {\it {Baryogenesis, domain walls and the role of gravity}},  {\em Phys. Lett. B} {\bf 309} (1993) 258--263, [\href{http://arxiv.org/abs/hep-ph/9304203}{{\tt hep-ph/9304203}}].

\bibitem{Zhang:2023nrs}
Z.~Zhang, C.~Cai, Y.-H. Su, S.~Wang, Z.-H. Yu, and H.-H. Zhang, {\it {Nano-Hertz gravitational waves from collapsing domain walls associated with freeze-in dark matter in light of pulsar timing array observations}},  {\em Phys. Rev. D} {\bf 108} (2023), no.~9 095037, [\href{http://arxiv.org/abs/2307.11495}{{\tt arXiv:2307.11495}}].

\bibitem{Zeng:2025zjp}
Q.-Q. Zeng, X.~He, Z.-H. Yu, and J.~Zheng, {\it {Collapsing domain walls with Z2-violating coupling to thermalized fermions and their impact on gravitational wave detections}},  {\em Phys. Rev. D} {\bf 111} (2025), no.~11 115017, [\href{http://arxiv.org/abs/2501.10059}{{\tt arXiv:2501.10059}}].

\bibitem{Casas:2001sr}
J.~Casas and A.~Ibarra, {\it {Oscillating neutrinos and $\mu \to e, \gamma$}},  {\em Nucl. Phys. B} {\bf 618} (2001) 171--204, [\href{http://arxiv.org/abs/hep-ph/0103065}{{\tt hep-ph/0103065}}].

\bibitem{Ibarra:2003up}
A.~Ibarra and G.~G. Ross, {\it {Neutrino phenomenology: The Case of two right-handed neutrinos}},  {\em Phys. Lett. B} {\bf 591} (2004) 285--296, [\href{http://arxiv.org/abs/hep-ph/0312138}{{\tt hep-ph/0312138}}].

\bibitem{Esteban:2024eli}
I.~Esteban, M.~C. Gonzalez-Garcia, M.~Maltoni, I.~Martinez-Soler, J.~P. Pinheiro, and T.~Schwetz, {\it {NuFit-6.0: updated global analysis of three-flavor neutrino oscillations}},  {\em JHEP} {\bf 12} (2024) 216, [\href{http://arxiv.org/abs/2410.05380}{{\tt arXiv:2410.05380}}].

\bibitem{Kibble:1976sj}
T.~W.~B. Kibble, {\it {Topology of Cosmic Domains and Strings}},  {\em J. Phys. A} {\bf 9} (1976) 1387--1398.

\bibitem{Coleman:1973jx}
S.~R. Coleman and E.~J. Weinberg, {\it {Radiative Corrections as the Origin of Spontaneous Symmetry Breaking}},  {\em Phys. Rev. D} {\bf 7} (1973) 1888--1910.

\bibitem{Dolan:1973qd}
L.~Dolan and R.~Jackiw, {\it {Symmetry Behavior at Finite Temperature}},  {\em Phys. Rev. D} {\bf 9} (1974) 3320--3341.

\bibitem{Quiros:1999jp}
M.~Quiros, {\it {Finite temperature field theory and phase transitions}},  in {\em {ICTP Summer School in High-Energy Physics and Cosmology}}, pp.~187--259, 1, 1999.
\newblock \href{http://arxiv.org/abs/hep-ph/9901312}{{\tt hep-ph/9901312}}.

\bibitem{Bai:2023cqj}
Y.~Bai, T.-K. Chen, and M.~Korwar, {\it {QCD-collapsed domain walls: QCD phase transition and gravitational wave spectroscopy}},  {\em JHEP} {\bf 12} (2023) 194, [\href{http://arxiv.org/abs/2306.17160}{{\tt arXiv:2306.17160}}].

\bibitem{Nakayama:2016gxi}
K.~Nakayama, F.~Takahashi, and N.~Yokozaki, {\it {Gravitational waves from domain walls and their implications}},  {\em Phys. Lett. B} {\bf 770} (2017) 500--506, [\href{http://arxiv.org/abs/1612.08327}{{\tt arXiv:1612.08327}}].

\bibitem{Galtsov:2017udh}
D.~Gal'tsov, E.~Melkumova, and P.~Spirin, {\it {Piercing of domain walls: new mechanism of gravitational radiation}},  {\em JHEP} {\bf 01} (2018) 120, [\href{http://arxiv.org/abs/1711.01114}{{\tt arXiv:1711.01114}}].

\bibitem{Babichev:2021uvl}
E.~Babichev, D.~Gorbunov, S.~Ramazanov, and A.~Vikman, {\it {Gravitational shine of dark domain walls}},  {\em JCAP} {\bf 04} (2022), no.~04 028, [\href{http://arxiv.org/abs/2112.12608}{{\tt arXiv:2112.12608}}].

\bibitem{Hiramatsu:2013qaa}
T.~Hiramatsu, M.~Kawasaki, and K.~Saikawa, {\it {On the estimation of gravitational wave spectrum from cosmic domain walls}},  {\em JCAP} {\bf 02} (2014) 031, [\href{http://arxiv.org/abs/1309.5001}{{\tt arXiv:1309.5001}}].

\bibitem{Hiramatsu:2012sc}
T.~Hiramatsu, M.~Kawasaki, K.~Saikawa, and T.~Sekiguchi, {\it {Axion cosmology with long-lived domain walls}},  {\em JCAP} {\bf 01} (2013) 001, [\href{http://arxiv.org/abs/1207.3166}{{\tt arXiv:1207.3166}}].

\bibitem{Kitajima:2023cek}
N.~Kitajima, J.~Lee, K.~Murai, F.~Takahashi, and W.~Yin, {\it {Gravitational waves from domain wall collapse, and application to nanohertz signals with QCD-coupled axions}},  {\em Phys. Lett. B} {\bf 851} (2024) 138586, [\href{http://arxiv.org/abs/2306.17146}{{\tt arXiv:2306.17146}}].

\bibitem{Notari:2025kqq}
A.~Notari, F.~Rompineve, and F.~Torrenti, {\it {The spectrum of gravitational waves from annihilating domain walls}},  {\em JCAP} {\bf 07} (2025) 049, [\href{http://arxiv.org/abs/2504.03636}{{\tt arXiv:2504.03636}}].

\bibitem{Kadota:2015dza}
K.~Kadota, M.~Kawasaki, and K.~Saikawa, {\it {Gravitational waves from domain walls in the next-to-minimal supersymmetric standard model}},  {\em JCAP} {\bf 10} (2015) 041, [\href{http://arxiv.org/abs/1503.06998}{{\tt arXiv:1503.06998}}].

\bibitem{Chen:2020wvu}
N.~Chen, T.~Li, and Y.~Wu, {\it {The gravitational waves from the collapsing domain walls in the complex singlet model}},  {\em JHEP} {\bf 08} (2020) 117, [\href{http://arxiv.org/abs/2004.10148}{{\tt arXiv:2004.10148}}].

\bibitem{Dankovsky:2024zvs}
I.~Dankovsky, E.~Babichev, D.~Gorbunov, S.~Ramazanov, and A.~Vikman, {\it {Revisiting evolution of domain walls and their gravitational radiation with CosmoLattice}},  {\em JCAP} {\bf 09} (2024) 047, [\href{http://arxiv.org/abs/2406.17053}{{\tt arXiv:2406.17053}}].

\bibitem{Blasi:2025tmn}
S.~Blasi, A.~Mariotti, A.~Rase, and M.~Vanvlasselaer, {\it {Domain walls in the scaling regime: Equal Time Correlator and Gravitational Waves}},  \href{http://arxiv.org/abs/2511.16649}{{\tt arXiv:2511.16649}}.

\bibitem{Caprini:2019egz}
C.~Caprini {\em et~al.}, {\it {Detecting gravitational waves from cosmological phase transitions with LISA: an update}},  {\em JCAP} {\bf 03} (2020) 024, [\href{http://arxiv.org/abs/1910.13125}{{\tt arXiv:1910.13125}}].

\bibitem{NANOGrav:2023hvm}
{\bf NANOGrav} Collaboration, A.~Afzal {\em et~al.}, {\it {The NANOGrav 15 yr Data Set: Search for Signals from New Physics}},  {\em Astrophys. J. Lett.} {\bf 951} (2023), no.~1 L11, [\href{http://arxiv.org/abs/2306.16219}{{\tt arXiv:2306.16219}}].

\bibitem{Domenech:2020ssp}
G.~Dom\`enech, C.~Lin, and M.~Sasaki, {\it {Gravitational wave constraints on the primordial black hole dominated early universe}},  {\em JCAP} {\bf 04} (2021) 062, [\href{http://arxiv.org/abs/2012.08151}{{\tt arXiv:2012.08151}}]. [Erratum: JCAP 11, E01 (2021)].

\bibitem{Yagi:2011wg}
K.~Yagi and N.~Seto, {\it {Detector configuration of DECIGO/BBO and identification of cosmological neutron-star binaries}},  {\em Phys. Rev. D} {\bf 83} (2011) 044011, [\href{http://arxiv.org/abs/1101.3940}{{\tt arXiv:1101.3940}}]. [Erratum: Phys.Rev.D 95, 109901 (2017)].

\bibitem{2017arXiv170200786A}
{\bf LISA} Collaboration, P.~Amaro-Seoane~et al, {\it {Laser Interferometer Space Antenna}},  {\em arXiv e-prints} (Feb., 2017) arXiv:1702.00786, [\href{http://arxiv.org/abs/1702.00786}{{\tt arXiv:1702.00786}}].

\bibitem{Punturo_2010}
{\bf ET Collaboration} Collaboration, M.~Punturo~et al, {\it The einstein telescope: a third-generation gravitational wave observatory},  {\em Classical and Quantum Gravity} {\bf 27} (sep, 2010) 194002.

\bibitem{Garcia-Bellido:2021zgu}
J.~Garcia-Bellido, H.~Murayama, and G.~White, {\it {Exploring the Early Universe with Gaia and THEIA}},  \href{http://arxiv.org/abs/2104.04778}{{\tt arXiv:2104.04778}}.

\bibitem{Sesana:2019vho}
A.~Sesana {\em et~al.}, {\it {Unveiling the gravitational universe at $\mu$-Hz frequencies}},  {\em Exper. Astron.} {\bf 51} (2021), no.~3 1333--1383, [\href{http://arxiv.org/abs/1908.11391}{{\tt arXiv:1908.11391}}].

\bibitem{Weltman:2018zrl}
A.~Weltman {\em et~al.}, {\it {Fundamental physics with the Square Kilometre Array}},  {\em Publ. Astron. Soc. Austral.} {\bf 37} (2020) e002, [\href{http://arxiv.org/abs/1810.02680}{{\tt arXiv:1810.02680}}].

\bibitem{NANOGrav:2023gor}
{\bf NANOGrav} Collaboration, G.~Agazie {\em et~al.}, {\it {The NANOGrav 15 yr Data Set: Evidence for a Gravitational-wave Background}},  {\em Astrophys. J. Lett.} {\bf 951} (2023), no.~1 L8, [\href{http://arxiv.org/abs/2306.16213}{{\tt arXiv:2306.16213}}].

\bibitem{Rosado:2011kv}
P.~A. Rosado, {\it {Gravitational wave background from binary systems}},  {\em Phys. Rev. D} {\bf 84} (2011) 084004, [\href{http://arxiv.org/abs/1106.5795}{{\tt arXiv:1106.5795}}].

\bibitem{Baldes:2023rqv}
I.~Baldes and M.~O. Olea-Romacho, {\it {Primordial black holes as dark matter: Interferometric tests of phase transition origin}},  \href{http://arxiv.org/abs/2307.11639}{{\tt arXiv:2307.11639}}.

\bibitem{Abazajian:2019eic}
K.~Abazajian {\em et~al.}, {\it {CMB-S4 Science Case, Reference Design, and Project Plan}},  \href{http://arxiv.org/abs/1907.04473}{{\tt arXiv:1907.04473}}.

\bibitem{CMB-HD:2022bsz}
{\bf CMB-HD} Collaboration, S.~Aiola {\em et~al.}, {\it {Snowmass2021 CMB-HD White Paper}},  \href{http://arxiv.org/abs/2203.05728}{{\tt arXiv:2203.05728}}.

\bibitem{Giudice:2003jh}
G.~Giudice, A.~Notari, M.~Raidal, A.~Riotto, and A.~Strumia, {\it {Towards a complete theory of thermal leptogenesis in the SM and MSSM}},  {\em Nucl. Phys. B} {\bf 685} (2004) 89--149, [\href{http://arxiv.org/abs/hep-ph/0310123}{{\tt hep-ph/0310123}}].

\bibitem{Hahn-Woernle:2009jyb}
F.~Hahn-Woernle, M.~Plumacher, and Y.~Y.~Y. Wong, {\it {Full Boltzmann equations for leptogenesis including scattering}},  {\em JCAP} {\bf 08} (2009) 028, [\href{http://arxiv.org/abs/0907.0205}{{\tt arXiv:0907.0205}}].

\bibitem{Harvey:1990qw}
J.~A. Harvey and M.~S. Turner, {\it {Cosmological Baryon and Lepton Number in the Presence of Electroweak Fermion Number Violation}},  {\em Phys. Rev. D} {\bf 42} (1990) 3344--3349.

\bibitem{Kawasaki:2004rx}
M.~Kawasaki and F.~Takahashi, {\it {Late-time entropy production due to the decay of domain walls}},  {\em Phys. Lett. B} {\bf 618} (2005) 1--6, [\href{http://arxiv.org/abs/hep-ph/0410158}{{\tt hep-ph/0410158}}].

\bibitem{Hattori:2015xla}
H.~Hattori, T.~Kobayashi, N.~Omoto, and O.~Seto, {\it {Entropy production by domain wall decay in the NMSSM}},  {\em Phys. Rev. D} {\bf 92} (2015), no.~10 103518, [\href{http://arxiv.org/abs/1510.03595}{{\tt arXiv:1510.03595}}].

\bibitem{Hong:2025piv}
S.~Hong, S.~M. Lee, and Q.~Liang, {\it {Gravitational Wave with Domain Wall Dominance}},  \href{http://arxiv.org/abs/2504.02462}{{\tt arXiv:2504.02462}}.

\bibitem{Borah:2017qdu}
D.~Borah, M.~K. Das, and A.~Mukherjee, {\it {Common origin of nonzero $\theta_{13}$ and baryon asymmetry of the Universe in a TeV scale seesaw model with $A_4$ flavor symmetry}},  {\em Phys. Rev. D} {\bf 97} (2018), no.~11 115009, [\href{http://arxiv.org/abs/1711.02445}{{\tt arXiv:1711.02445}}].

\bibitem{Barr:1992qq}
S.~M. Barr and D.~Seckel, {\it {Planck scale corrections to axion models}},  {\em Phys. Rev. D} {\bf 46} (1992) 539--549.

\end{thebibliography}\endgroup

\end{document}